\documentclass[sigconf, nonacm, authorversion,screen]{acmart}
\AtBeginDocument{%
  }

\setcopyright{acmlicensed}
\copyrightyear{2027}
\acmYear{2027}
\acmDOI{XXXXXXX.XXXXXXX}
\acmConference[WWW '27]{the ACM Web Conference 2027}{May 10--14, 2027}{Dublin, Ireland}
\acmISBN{978-1-4503-XXXX-X/2027/05}

\usepackage{mathtools}
\usepackage{multirow}
\usepackage{algorithm}
\usepackage{algpseudocode}
\usepackage{enumitem}

\theoremstyle{acmplain}
\newtheorem{theorem}{Theorem}
\newtheorem{lemma}{Lemma}

\newtheorem{proposition}{Proposition}
\newtheorem{assumption}{Assumption}
\theoremstyle{acmdefinition}
\newtheorem{definition}{Definition}

\usepackage{refcount}
\usepackage[capitalize,nameinlink]{cleveref}
\crefname{problem}{Problem}{Problems}
\Crefname{problem}{Problem}{Problems}
\crefname{definition}{Definition}{Definitions}
\Crefname{definition}{Definition}{Definitions}
\crefname{assumption}{Assumption}{Assumptions}
\Crefname{assumption}{Assumption}{Assumptions}
\crefname{algorithm}{Algorithm}{Algorithms}
\Crefname{algorithm}{Algorithm}{Algorithms}

\newcommand{\R}{\mathbb{R}}
\newcommand{\E}{\mathbb{E}}

\renewcommand{\hat}{\widehat}
\newcommand{\xhdr}[1]{\noindent\textbf{#1}}
\DeclareMathOperator{\diag}{diag}

\DeclareMathOperator{\rank}{rank}
\DeclareMathOperator*{\argmin}{arg\,min}
\DeclareMathOperator{\col}{col}
\renewcommand{\tilde}{\widetilde}

\begin{document}

\title{Discovering Topics from Linear Autoencoders}

\author{Jutiyakhun Srisoparp}
\affiliation{%
  \institution{Chiang Mai University}
  \city{Chiang Mai}
  \country{Thailand}
}
\email{jutiyakhun\_s@cmu.ac.th}

\author{Donlapark Ponnoprat}
\correspondingauthor
\affiliation{%
  \institution{Chiang Mai University}
  \city{Chiang Mai}
  \country{Thailand}
}
\email{donlapark.p@cmu.ac.th}

\renewcommand{\shortauthors}{Jutiyakhun Srisoparp and Donlapark Ponnoprat}

\begin{abstract}
Linear recommenders and topic models are similar in that they implicitly learn from co-occurrences of items. However, there has been little study of how these two model families are related. In this work, we show that, under a latent-factor model of the user-item interactions, the latent topics can be discovered from the weight matrix of a closed-form linear autoencoder. These topics provide a basis for interpreting the recommendations, and can be obtained directly from the trained recommender without additional training. Building on our analysis, we further propose novel item and user normalizations that prevent popular items and very active users from dominating the topics. Our experiments on six public datasets demonstrate that our method achieves significantly higher topic quality than every dedicated topic model and matches or exceeds the best linear autoencoder on all six datasets, while its recommendation performance is competitive with the best tuned recommender.
\end{abstract}




\maketitle

\section{Introduction} \label{sec:intro}

A linear recommender and a topic model can both be fitted from the same user-item interactions. Specifically, given a binary user-item matrix $X\in\{0,1\}^{n\times m}$ for $n$ users and $m$ items, a linear recommender scores item $i$ for user $u$ by $(XB)_{ui}$, where $B\in\R^{m\times m}$ is an \emph{item-item weight matrix} fitted from $X$ \citep{ning2011slim,steck2019ease}. A topic model instead summarizes the $m$ items by $K$ latent groups, called \emph{topics}, where each topic is represented by a discrete distribution over the items \citep{blei2003lda,dieng2020etm}.

There have been several techniques that incorporate topic models in a recommender system \citep{wang2011ctr,mcauley2013hft,gopalan2014ctpf}. However, these techniques require textual descriptions or reviews of the items as well as a jointly fitted topic model. We ask instead: can the topics be estimated from the weight matrix $B$ of a linear recommender, without item texts and without training a separate topic model? To answer this question, we focus on the family of \emph{linear autoencoders} \citep{ning2011slim,steck2019ease,steck2020dlae,moon2023rlae}, which fit $B$ so that $XB$ approximates $X$, subject to $B_{ii}=0$ for every item $i$, so that no item contributes to its own score. Specifically, we study the linear autoencoders whose solutions are available in closed form. The best known of such autoencoders is EASE \citep{steck2019ease}, whose weight matrix $B$ is computed from the co-occurrence counts of the items. EASE has attracted interest due to its simplicity and strong recommendation performance, giving rise to several variants \citep{steck2020dlae,moon2023rlae,park2025dan}.

Despite their success, no prior work has studied whether these linear autoencoders can recover the latent topics when the interaction data is assumed to follow a topic model. In this paper, we answer this question and use the answer to build a linear autoencoder that provides both recommendations and topics.

Our contributions are summarized as follows.
\begin{itemize}[leftmargin=5mm]
  \item \textbf{Topics in linear autoencoders.} We show that, when the interactions follow a latent-factor model, the topics can be estimated from the weight matrix of a closed-form linear autoencoder.
  \item \textbf{Item and user normalizations.} We propose \emph{Topic-Enabled Autoencoders} (TEA), which apply novel item and user normalizations to the item co-occurrence matrix. Unlike previous work \citep{park2025dan}, our user normalization does not require any hyperparameter tuning.
  \item \textbf{Empirical results.} We demonstrate that TEA achieves significantly higher topic quality than every dedicated topic model and matches or exceeds the best linear autoencoder on all six public datasets, while its recommendation performance is competitive with the best tuned recommender.
\end{itemize}

\paragraph{Notations.} For any matrix $A$, $A_{ij}$ denotes the $(i,j)$-entry of $A$. For any sets $I,J$ of positive integers, $A_{I,J}$ denotes the submatrix of $A$ with rows in $I$ and columns in $J$. Similarly, $A_{i,J}$ and $A_{I,i}$ denote the corresponding parts of the $i$-th row and the $i$-th column of $A$. For a square matrix $A$, we write $\diag(A)$ for the diagonal matrix whose $i$-th diagonal entry is $A_{ii}$. For a vector $a$, we write $\diag(a)$ for the diagonal matrix whose $i$-th diagonal entry is $a_i$. For a (possibly non-square) matrix $B$, we denote by $\col(B)$ the column space of $B$ and $\rank(B)$ the rank of $B$. We write $\mathbf{1}\in\R^m$ for the all-ones vector.

\section{Setup}
\label{sec:setup}

We observe $n$ users interacting with $m$ items as a binary matrix $X\in\{0,1\}^{n\times m}$, where $X_{ui}=1$ if user $u$ has interacted with item $i$ and $X_{ui}=0$ otherwise. Throughout this paper, users are indexed by $u$ and items are indexed by $i,j$. We write $x_u\in\{0,1\}^{m}$ for the $u$-th row of $X$, as a column vector. The \emph{co-occurrence matrix} of the interactions is the following \emph{Gram matrix}:
\begin{equation*}
  G\;=\;X^\top X\;=\;\sum_{u=1}^{n}x_ux_u^\top,
  \qquad
  G_{ij}\;=\;\sum_{u=1}^{n}X_{ui}X_{uj},
\end{equation*}
where $G_{ij}$ is the number of users who interacted with both item $i$ and item $j$. In particular, the diagonal entry $g_i=G_{ii}$ is the number of users who interacted with item $i$, which we refer to as the \emph{popularity} of item $i$.

\subsection{Closed-form linear autoencoders} \label{sec:closedform}
Given an interaction matrix $X\in\{0,1\}^{n\times m}$, an \emph{item-based recommender} scores the items by an item-item \emph{weight matrix} $B\in\R^{m\times m}$. In regression-based recommenders, the $j$-th column of $B$ contains the coefficients of a linear regression of the interactions of item $j$ on those of the other items. The score of item $i$ for user $u$ is $(XB)_{ui}=\sum_j X_{uj}B_{ji}$, that is, the sum of the weights $B_{ji}$
over the items $j$ in $u$'s history. In this work, we focus on item-based recommenders whose weight matrix $B$ has a closed form. The best known of these is EASE \citep{steck2019ease}, which solves the following problem, with a constraint $\diag(B)=\mathbf{0}$ so that no item contributes to its own score:
\begin{equation}\label{eq:ease}
  \tilde B\;=\;\argmin_{B:\,\diag(B)=\mathbf{0}}\;\lVert X-XB\rVert_F^2+\lambda\lVert B\rVert_F^2 ,
\end{equation}
where $\lVert\cdot\rVert_F$ is the Frobenius norm and $\lambda>0$ is a regularization parameter. For a square matrix $A$, let $\mathcal H(A)=A-\diag(A)$ denote the off-diagonal matrix of $A$. With the inverse $P\coloneqq(G+\lambda I)^{-1}$ of the regularized co-occurrence matrix, the solution of \eqref{eq:ease} is
\begin{equation}\label{eq:easeclosed}
  \tilde B\;=\;I-P\diag(P)^{-1}\;=\;-\mathcal H(P)\,\diag(P)^{-1}.
\end{equation}
Here, the right multiplication by $\diag(P)^{-1}$ multiplies the $i$-th column of $-\mathcal H(P)$ by $P_{ii}^{-1}$.

Motivated by \eqref{eq:easeclosed}, we consider a broader class of weight matrices for recommender systems.
\begin{definition}[GLAE]\label{def:family}
A \emph{generalized linear autoencoder (GLAE)} is an item-item weight matrix
of the form:
\begin{equation}\label{eq:glae}
  B_0\;=\;-\mathcal H\!\big((G_0+\Lambda)^{-1}\big)\,\Sigma,
\end{equation}
where $G_0\in\R^{m\times m}$ is symmetric and positive semidefinite, and $\Lambda,\Sigma\in\R^{m\times m}$ are diagonal matrices with positive diagonal entries.
\end{definition}
By \eqref{eq:easeclosed}, EASE is the GLAE \eqref{eq:glae} with $G_0=G$, $\Lambda=\lambda I$, and $\Sigma=\diag(P)^{-1}$.

\subsection{Latent-factor models of user-item interactions}\label{sec:topicmodels}

We now propose a latent-factor model of $X$ whose latent factors define the topics. Since we want the topics to be diverse, the latent factors should not be dominated by popular items or very active users. To this end, the model places item popularity and user activity in per-item and per-user weights, separate from the latent factors. The following definition formalizes this idea.

\begin{definition}[wLFM] \label{def:model}
We say that the interaction matrix $X$ \emph{follows} the \emph{weighted latent factor model (wLFM)} if there exist a matrix $\beta \in \R^{m \times K}_{+}$, user factors $f_1,\dots,f_n\in\R_+^{K}$, and scalars $w^I_1,\dots,w^I_m>0$ and $w^U_1,\dots,w^U_n>0$ such that, for every user $u$ and item $i$, $X_{ui}$ is a random variable with mean
\begin{equation}\label{eq:linmean}
  \E[X_{ui}\mid w^I_i, w^U_u, \beta_{i,:},f_u]\;=\; w^I_i w^U_u \langle\beta_{i,:},f_u\rangle.
\end{equation}
In addition, for every user $u$ and any two items $i \not= j$, $X_{ui}$ and $X_{uj}$ are conditionally independent given $w^I_i,w^I_j,w^U_u,\beta_{i,:}$, $\beta_{j,:}$, and $f_u$. We call the $i$-th row $\beta_{i,:}$ of $\beta$ the \emph{item factor} of item $i$ and $f_u$ the \emph{user factor} of user $u$.
\end{definition}
For brevity, we will omit the conditioning variables for the rest of the paper and write $\E[X_{ui}]\;=\; w^I_i w^U_u\langle\beta_{i,:},f_u\rangle$. We note that the condition in \eqref{eq:linmean} with unit weights is satisfied by the factor analysis model \citep{Everitt1984,roweis1999unifying} and the aspect Bernoulli model \citep{bingham2009aspect}, both of which have been applied to recommender systems \citep{marlin2003urp,srebro2004learning}.

\subsection{User and item normalizations}\label{sec:normalization}

To remove the weights $w^I_i$ and $w^U_u$ from the mean \eqref{eq:linmean}, we normalize $X$ by the diagonal matrices of these weights:
\begin{align}\label{eq:AB}
  \hat{X} \coloneqq D^{-1}_U X D^{-1}_I,\qquad \hat{G} \coloneqq \hat{X}^\top \hat{X} = D^{-1}_I X^\top D^{-2}_U X D^{-1}_I,
\end{align}
where $D_I=\diag(w^I_1,\dots,w^I_m)$ and $D_U=\diag(w^U_1,\dots,w^U_n)$. It then follows that $\E[\hat{X}_{ui}] = \langle\beta_{i,:},f_u\rangle$ for every user $u$ and item $i$.

We will focus on the normalized matrix $\hat{G}$ for the rest of the paper. First, we show that $\E[\hat{G}]$ is the sum of a positive semidefinite matrix of rank at most $K$ and a diagonal matrix.
\begin{lemma}\label{lem:lowrank}
If $X$ follows the wLFM, the expected normalized co-occurrence matrix decomposes as
\begin{equation*}
\E[\hat{G}] = L + D, \ \ L = \beta \Phi \beta^\top, \ \ \Phi = \sum_{u=1}^n f_u f_u^\top, 
\end{equation*}
and $D$ is the diagonal matrix with $D_{ii} = \sum_{u=1}^n \mathrm{Var}(\hat X_{ui})$. In particular, $L \succeq 0$ and $\rank(L) \le K$.
\end{lemma}

In view of \cref{lem:lowrank}, we make the following assumption on the expected normalized co-occurrence matrix:
\begin{assumption}\label{ass:lrd}
The normalized co-occurrence matrix $\hat G$ in \eqref{eq:AB} satisfies $G^\star\coloneqq\E[\hat G]=L+D$, where $L=\beta \Phi \beta^\top$, $\beta\in\R_+^{m\times K}$, $D$ is a diagonal matrix with positive diagonal entries, $\rank(\beta)=K$ and $\Phi \succ 0$.
\end{assumption}
The most important components of this assumption are the matrices $\beta$ and $D$, because our definition of topics will be in terms of these two matrices.

\subsection{Topic models from latent-factor models}\label{sec:topics}

Each topic is defined by one column of $\beta$ and the diagonal matrix $D$.
\begin{definition}[Topic]
Let $\beta$ and $D$ be as in \cref{ass:lrd}. For $k=1,\ldots,K$, the $k$\emph{-th topic} is the solution of the constrained maximization of the signal-to-noise ratio
\begin{equation}\label{eq:obj}
\max_{v\in\mathbb{R}^m}
\left\{
\frac{(v^\top\beta_{:,k})^2}{v^\top D v}
:
\sum_{i=1}^m v_i=1
\right\}.
\end{equation}
\end{definition}
In other words, the $k$-th topic is the allocation of the items that maximizes the mean interaction along the $k$-th latent factor, while ensuring that each user's behavior is consistent (i.e. has low variance) under this allocation.

The maximization problem \eqref{eq:obj} has the following closed-form solution.
\begin{proposition}\label{prop:topicdir}
Suppose that \cref{ass:lrd} holds. Then, for every $k$, the maximization problem \eqref{eq:obj} has the unique solution
\[
t_k=\frac{D^{-1}\beta_{:,k}}{\mathbf{1}^\top D^{-1}\beta_{:,k}}.
\]
\end{proposition}
Thus, the $k$-th topic is the $k$-th column of $D^{-1}\beta$, normalized so that its entries sum to one. Since the entries of $D^{-1}\beta$ are non-negative, $t_k$ is a probability distribution over the items. Moreover, the order of the topics carries no meaning. This leads to the following notion of a topic model.

\begin{definition}[topic model]\label{def:ktopic}
Let $t_1,\dots,t_K$ be the topics of \cref{prop:topicdir}. The \emph{topic model} is the equivalence class of the matrix $\mathcal{T}_K=[t_1,\dots,t_K]\in\R_+^{m\times K}$ under permutation of its columns.
\end{definition}

\subsection{Topic recovery from a GLAE}\label{sec:problem}

We now tie everything together and state our goal. Motivated by the decomposition of $\E[\hat G]$ in \cref{lem:lowrank}, we focus on the GLAE \eqref{eq:glae} with $G_0 = \hat{G}$. Under \cref{ass:lrd}, our goal is to recover the topic model of \cref{def:ktopic} from this GLAE in two steps:

\paragraph{Step 1: Choosing the normalizations.} We narrow down our choice of GLAE by (1) choosing the item weights $w^I_i$ and the user weights $w^U_u$, which determine the normalized co-occurrence matrix $\hat G$ in \eqref{eq:AB}, and (2) choosing the regularization matrix $\Lambda$ in \eqref{eq:glae}. Our choices of these weights and of $\Lambda$ are in \cref{sec:usernorm}.

\paragraph{Step 2: Recovering the topics.} We analyze whether the topic model $\mathcal T_K$ can be recovered from the weight matrix $B_0$ of \eqref{eq:glae} when $G_0=\hat G$, the normalized co-occurrence matrix. In \Cref{sec:identification}, we show that the GLAE with $G_0=\E[\hat G]$ determines the matrix $(D+\Lambda)^{-1}\beta$ up to permutation and positive column scaling. We normalize the columns of this matrix to estimate $\mathcal T_K$, where $\Lambda$ regularizes the inverse of $D$. In \Cref{sec:observed}, we analyze the GLAE with the observed matrix $G_0=\hat G$ and bound how far its eigenspace is from $\col((D+\Lambda)^{-1}\beta)$.

\section{Topic recovery from linear autoencoders}
\label{sec:identification}

We first study the GLAE \eqref{eq:glae} with $G_0=\E[\hat G]$ and with a regularization matrix $\Lambda$. Throughout this section, we assume that \cref{ass:lrd} holds, with $\beta$, $\Phi$, and $D$ as in that assumption. For brevity, we write:
\begin{equation}\label{eq:gauge}
\begin{aligned}
\beta_\Lambda&=(D+\Lambda)^{-1}\beta, &\qquad M&=\Phi^{-1}+\beta^\top\beta_\Lambda,\\
P^\star&=(G^\star+\Lambda)^{-1}, & S^\star&=\beta_\Lambda M^{-1}\beta_\Lambda^\top.
\end{aligned}
\end{equation}
When $\Lambda=\mathbf{0}$, $\beta_\Lambda=D^{-1}\beta$, so normalizing the columns of $\beta_\Lambda$ to sum to one gives the matrix $\mathcal T_K$ of \cref{def:ktopic}. However, some diagonal entries of $D$ can be close to zero, so that the corresponding entries of $D^{-1}$ are large. The regularization matrix $\Lambda$ keeps the diagonal entries of $(D+\Lambda)^{-1}$ bounded.

To identify $\col(\beta_\Lambda)$ from $G^\star$, we make the following assumption, which is the condition of \citet[Theorem~5.1]{anderson1956statistical} for identification in factor analysis.
\begin{assumption}\label{ass:ident}
For every item $i$, there are disjoint sets $I,J\subset\{1,\dots,m\}\setminus\{i\}$ of $K$ items each such that the submatrices $\beta_{I,:}$ and $\beta_{J,:}$ are invertible.
\end{assumption}
Note that \Cref{ass:ident} requires $m\ge 2K+1$. The next two results show that, under these assumptions, the column space of $\beta_\Lambda$ can be obtained from the GLAE with $G_0=G^\star$.

\begin{lemma}\label{lem:recovery}
Suppose that \cref{ass:lrd} holds. Then, $M\succ0$ and
\[
-\mathcal H(P^\star)=\mathcal H(S^\star).
\]
Moreover, $S^\star$ is positive semidefinite of rank $K$ with column space $\col(\beta_\Lambda)$.
\end{lemma}
The proof is in \cref{app:recoveryproof}. By \cref{lem:recovery}, $G^\star$ determines $\col(\beta_\Lambda)$ once the diagonal entries of $S^\star$ are known. The following proposition shows that the off-diagonal entries of $P^\star$ determine these diagonal entries:
\begin{proposition}\label{prop:ident}
Suppose that \cref{ass:lrd,ass:ident} hold. Then $S^\star$ is the only matrix of rank at most $K$ whose off-diagonal matrix is $-\mathcal H(P^\star)$. Moreover, for every item $i$, with $I$ and $J$ as in \cref{ass:ident},
\begin{equation}\label{eq:sstar}
S^\star_{ii}=-P^\star_{i,J}\,\big(P^\star_{I,J}\big)^{-1}\,P^\star_{I,i}.
\end{equation}
Finally, for disjoint sets $I,J\subset\{1,\dots,m\}$ of $K$ items each, the submatrix $P^\star_{I,J}$ is invertible if and only if $\beta_{I,:}$ and $\beta_{J,:}$ are invertible.
\end{proposition}
The proof is in \cref{app:identproof}. Since $i\notin I\cup J$ and $I\cap J=\emptyset$, the calculation in \eqref{eq:sstar} uses only off-diagonal entries of $P^\star$. As a consequence, if we have access to $G^\star$, $\col(\beta_\Lambda)$ can be recovered in two steps. First, for every item $i$, we choose disjoint sets $I$ and $J$ of $K$ items not containing $i$ such that $P^\star_{I,J}$ is invertible. Second, we compute $S^\star_{ii}$ using \eqref{eq:sstar}, which, combined with the off-diagonal entries of $-\mathcal H(P^\star)$, gives $S^\star$, whose column space is $\col(\beta_\Lambda)$ by \cref{lem:recovery}.

To obtain $\beta_\Lambda$ from $\col(\beta_\Lambda)$, we require a condition that determines $\beta_\Lambda$ up to permutation and positive column scaling. For this purpose, we adopt the anchor condition of separable non-negative matrix factorization and of anchor-word topic models \citep{donoho2003nmf,gillis2014spa,arora2012topic}:

\begin{definition}\label{def:anchor}
Consider a matrix $U\in\R_+^{m\times K}$. An item $a_k$ is an \emph{anchor} of topic $k$ if $U_{a_k, k} > 0$ and $U_{a_k, k'} = 0$ for each $k' \neq k$. The matrix $U$ is \emph{separable} if each topic has an anchor and the $K$ anchors are distinct items.
\end{definition}

\begin{assumption}\label{ass:sep}
The matrix $\beta$ is separable.
\end{assumption}
We are now in a position to state the main identification result for $\beta_\Lambda$.
\begin{theorem}[uniqueness under separability]\label{thm:sep}
Suppose that \cref{ass:lrd,ass:sep} hold. Then $\beta_\Lambda$ is separable. Moreover, every separable non-negative basis of $\col(\beta_\Lambda)$ is equal to $\beta_\Lambda$ up to permutation and positive column scaling.
\end{theorem}

The proof is in \cref{app:sepproof}. There are several algorithms for obtaining $\beta_\Lambda$ from $\col(\beta_\Lambda)$, such as the successive projection algorithm (SPA) \citep{gillis2014spa} and its variant, the pseudo-point successive projection algorithm (ppSPA) \citep{jin2024ppspa}. The permutation and the column scaling in \cref{thm:sep} do not affect the topics. Indeed, we can normalize each column of $\beta_\Lambda$ to sum to one to remove the column scaling.

\section{Topic recovery from the observed matrix} \label{sec:observed}

In practice, we compute the GLAE \eqref{eq:glae} with $G_0=\hat G$, the observed matrix, instead of $G^\star$. The resulting GLAE is $\hat{B}=\hat A\,\Sigma$, where
\begin{equation}\label{eq:observedA}
\hat P=(\hat G+\Lambda)^{-1},\qquad \hat A=-\mathcal H(\hat P),
\end{equation}
and $\Sigma$ is a known diagonal matrix. In particular, $\hat P$ is a part of the calculation of $\hat B$, with which we will estimate the topic model.

However, the results in \cref{sec:identification} do not apply here since the off-diagonal entries of $\hat A$ need not be those of a rank-$K$ matrix. In particular, the formula of \eqref{eq:sstar} in \cref{prop:ident} can give different diagonal entries for different choices of $I$ and $J$. For this reason, unlike the method described in the previous section, we do not fill in the diagonal entries of $\hat A$. The following theorem shows that we can instead keep the diagonal entries of $\hat A$ as zero and estimate $\col(\beta_\Lambda)$ by an eigenspace of $\hat A$ itself. 

Let $\hat V\in\R^{m\times K}$ be a matrix whose columns are orthonormal eigenvectors of $\hat A$ associated with its $K$ largest eigenvalues. We measure how far $\col(\hat V)$ is from $\col(\beta_\Lambda)$ by calculating $\min_{C}\lVert S^\star-\hat VC\rVert$ where the minimum is taken over all $\R^{K\times m}$ matrices $C$, and $\lVert\cdot\rVert$ is the spectral norm.

The following theorem provides a bound for the error of estimation of $\col(\beta_\Lambda)$ using $\col(\hat{V})$: 

\begin{theorem}[error of the estimated subspace]\label{thm:subspace}
Suppose that \cref{ass:lrd} holds, and let $S^\star$ be as in \eqref{eq:gauge}. Let $\nu$ be the smallest diagonal entry of $D+\Lambda$, and let $\rho=\max_iL_{ii}/G^\star_{ii}$. Then, with $\hat V$ as above,
\begin{equation}\label{eq:subspacebound}
\min_{C\in\R^{K\times m}}\big\lVert S^\star-\hat VC\big\rVert\le\frac{5\,\lVert\hat G-G^\star\rVert}{\nu^2}+\frac{2\rho}{\nu}.
\end{equation}
\end{theorem}

The proof is in \cref{app:subspaceproof}. The two terms in \eqref{eq:subspacebound} correspond to the two ways in which $\hat A$ differs from $S^\star$. The first term is due to the difference between $\hat G$ and $G^\star$. The second term is the price of keeping the diagonal entries of $\hat A$ at zero instead of filling in those of $S^\star$ (which are not observed). This term arises from the fact that $S^\star$ is related to the latent factor matrix $L$, whose diagonal entries are at most a fraction $\rho$ of the diagonal entries of $G^\star=L+D$. Hence, this term is small when $D$ accounts for most of each diagonal entry of $G^\star$. In view of \cref{lem:lowrank}, $L_{ii}=\sum_u\E[\hat X_{ui}]^2$ and $D_{ii}=\sum_u\mathrm{Var}(\hat X_{ui})$, so this is the case when the interactions with each item are noisy relative to their means.

\section{Topic-Enabled Autoencoders (TEA)} \label{sec:usernorm}

We now introduce \emph{\textbf{T}opic-\textbf{E}nabled \textbf{A}utoencoders (\textbf{TEA})}, which is the GLAE \eqref{eq:glae} with $G_0=\hat G$ for a specific choice of the item weights $w^I_i$, the user weights $w^U_u$, and the regularization matrix $\Lambda$. The algorithm of TEA is summarized in \Cref{alg:tea}. Below, we explain each component of the algorithm.

\paragraph{Item normalization.} Popular items can lower the diversity of the topics. Indeed, a popular item co-occurs with many other items, so the item can appear among the highest-weight items of many topics. To reduce this effect of popularity, we propose an item normalization by setting the item weights to $w^I_i=g_i^{\alpha}$ in \eqref{eq:AB}, where $\alpha \in \R$ is a tuning parameter. A positive $\alpha$ down-weights the entries of popular items relative to those of less popular items, while a negative $\alpha$ up-weights them. For an item with $g_i=0$, that is, an item that no user interacted with, we set $w^I_i=1$ so that every item weight is positive.

\paragraph{User normalization.} Very active users can lower the accuracy of the estimated topics. Specifically, a very active user has many interactions, which add a large variance to $\hat G$. To reduce the effect of user activity on the topics, we propose a user normalization that chooses the user weights so that the variance of $\hat G$ is small relative to its off-diagonal entries.

In \eqref{eq:AB}, the user weights affect $\hat G$ only through the diagonal matrix $D_U^{-2}$. We write $D_U^{-2}=\diag(q_1,\dots,q_n)$ so that the weighted co-occurrence matrix is $\bar G=X^\top D_U^{-2}X=\sum_uq_ux_ux_u^\top$. Let $d_u=\sum_iX_{ui}$ be the number of items that user $u$ interacted with, and write $d=(d_1,\dots,d_n)$. The variance that user $u$ adds to $\bar G$ given $d$ depends on the matrix $A_u=\E[x_ux_u^\top\mid d]$, whose $(i,j)$ entry is the probability that user $u$ interacted with both items $i$ and $j$. The following theorem gives the weights $q_u$ that minimize a bound on the variance that the users add to $\bar G$, relative to the off-diagonal entries of $\bar G$.

\begin{theorem}[user weights]\label{thm:userweight}
Write $d=(d_1,\dots,d_n)$ and $A_u=\E[x_ux_u^\top\mid d]$. Let the weights $q_u\ge0$ depend only on $d$, let $Y_u=q_ux_ux_u^\top$, and write $\sigma_u=d_u(d_u-1)$ and $v_u=d_u\lVert A_u\rVert$. Then
\begin{align}
\E\big[\mathbf 1^\top\mathcal H(\bar G)\mathbf 1\mid d\big]&=\sum_{u=1}^nq_u\sigma_u,\label{eq:signal}\\
\Big\lVert\sum_{u=1}^n\mathrm{Var}(Y_u\mid d)\Big\rVert&\le\sum_{u=1}^nq_u^2v_u. \label{eq:noise}
\end{align}
Moreover, among the weights with $\sum_uq_u\sigma_u>0$, the ratio
\[
F(q)=\frac{\big(\sum_uq_u^2v_u\big)^{1/2}}{\sum_uq_u\sigma_u}
\]
is the smallest if and only if $q_u\propto(d_u-1)/\lVert A_u\rVert$ for every user $u$ with $d_u\ge1$.
\end{theorem}

The proof is in \cref{app:userweight}. The optimal weights depend on the matrices $A_u$, which are unknown. We therefore approximate $\lVert A_u\rVert$ by a function of $d_u$ via the following probabilistic perspective. Suppose that the $d_u$ interactions of user $u$ are independent draws from a distribution $\pi_u\in\R^m_+$ over the items. Then the expected number of $u$'s interactions with item $i$ is $d_u\pi_{ui}$. Similarly, for $i\neq j$, the expected number of ordered pairs of $u$'s interactions with items $i$ and $j$ is $d_u(d_u-1)\pi_{ui}\pi_{uj}$. Hence, when repeated draws of the same item are rare,
\[ A_u\;\approx\;d_u(d_u-1)\,\mathcal{H}(\pi_u\pi_u^\top)+d_u\diag(\pi_u).\]
By the triangle inequality, the spectral norm of the right-hand side is at most $\lVert\mathcal{H}(\pi_u\pi_u^\top)\rVert d_u^2+(\max_i\pi_{ui})d_u$. Here, high values of $\lVert \mathcal{H}(\pi_u\pi_u^\top) \rVert$ and $\max_i\pi_{ui}$ indicate that the user's interactions are concentrated only on a few items. But it is difficult to estimate $\pi_u$ from the user's binary interactions, so we instead posit the following approximation for $\lVert A_u\rVert$:
\[
  \lVert A_u\rVert\;\lesssim\; \lVert\mathcal{H}(\pi_u\pi_u^\top)\rVert d_u^2+(\max_i\pi_{ui})d_u \;\approx\; \hat{s}\,d_u^2+\hat{t}\,d_u,
\]
where $\hat s$ and $\hat t$ play the same roles as $\lVert \mathcal{H}(\pi_u\pi_u^\top)\rVert$ and $\max_i\pi_{ui}$, respectively. In particular, we choose $\hat s$ based on the pairs of distinct items: The pair $(i,j)$ accounts for a fraction of about $\pi_{ui}\pi_{uj}$ of the $d_u(d_u-1)$ ordered pairs of user $u$, so the matrix of these fractions is close to $\mathcal{H}(\pi_u\pi_u^\top)$. Accordingly, we choose $\hat s$ to be the largest eigenvalue of the average of these matrices over the users. Thus, we choose:
\begin{align*}
  \hat s = \lambda_1\Big(\frac1n\sum_{u:\,d_u\ge2}\frac{\mathcal{H}(x_ux_u^\top)}{d_u(d_u-1)}\Big), \quad \hat t = \max_ig_i/\sum_{i'}g_{i'},
\end{align*}
where $\lambda_1(\cdot)$ denotes the largest eigenvalue of the matrix. 

With this approximation, the optimal weights of \cref{thm:userweight} become $q_u\propto(d_u-1)/\big(d_u(d_u+\hat\tau)\big)$
\begin{equation}\label{eq:userweight}
  q_u \;=\; c\,\frac{d_u-1}{d_u\,(d_u+\hat\tau)},
\end{equation}
where $\hat\tau=\hat{t}/\hat{s}$, and $c>0$ is chosen so that $q_1,\dots,q_n$ sum to $n$ to maintain the original scale of $\hat G$. Since $\hat\tau$ is computed from $X$, the user weights require no tuning.

\paragraph{Matrix regularization.} We use the \emph{dropout regularization} proposed in \citep{steck2020dlae,moon2023rlae}:
\[
  \Lambda \;=\; \frac{p}{1-p}\,\diag(G)+\lambda I,
\]
where $p\in[0,1)$ is the dropout probability and $\lambda>0$ is the regularization parameter. With $\hat P$ of \eqref{eq:observedA}, the weight matrix of TEA is given by \eqref{eq:easeclosed} with $P$ replaced by $\hat P$:
\[
  \hat B \;=\; -\mathcal H(\hat P)\,\diag(\hat P)^{-1}.
\]

\paragraph{Topic estimation.} To extract the topics, we compute the matrix $\hat V$ of the $K$ leading eigenvectors of the matrix $\hat A=\hat B\diag(\hat P)$ in \eqref{eq:observedA}, whose column space estimates $\col(\beta_\Lambda)$ by \cref{thm:subspace}. To obtain $\hat\beta$ from the rows of $\hat V$, we apply the ppSPA algorithm \citep{jin2024ppspa}, which allows us to obtain $\beta_\Lambda$ from $\col(\beta_\Lambda)$ in \cref{sec:identification}. Finally, to follow the definition of a topic model in \cref{def:ktopic}, we normalize each column of $\hat\beta$ to sum to one and use the columns of $\hat\beta$ as the topics.

\begin{algorithm}[t]
\caption{TEA: the weight matrix and the topics}\label{alg:tea}
\begin{algorithmic}[1]
\Require Interaction matrix $X\in\{0,1\}^{n\times m}$, number of topics $K$, item exponent $\alpha$, dropout probability $p\in[0,1)$, regularization parameter $\lambda>0$
\Ensure Weight matrix $\hat B\in\R^{m\times m}$ and topics $\hat\beta\in\R_+^{m\times K}$, an estimate of the topic model $\mathcal T_K$
\State $G\gets X^\top X$, and $d_u\gets\sum_i X_{ui}$ for each user $u$
\State $\hat\tau\gets\dfrac{\max_iG_{ii}\big/\sum_{i'}G_{i'i'}}{\lambda_1\Big(\frac1n\sum_{u:\,d_u\ge2}\frac{x_ux_u^\top-\diag(x_u)}{d_u(d_u-1)}\Big)}$
\State $q_u\gets(d_u-1)\big/\big(d_u(d_u+\hat\tau)\big)$ for each user $u$
\State $q_u\gets q_u\big/\big(\frac1n\sum_{v}q_v\big)$ for each user $u$ 
\State $Q\gets\diag(q_1,\dots,q_n)$ \State $D_I\gets\diag(\max(G_{11},1)^{\alpha},\dots,\max(G_{mm},1)^{\alpha})$
\State $\hat G\gets D_I^{-1}X^\top QXD_I^{-1}$
\State $\Lambda\gets\frac{p}{1-p}\diag(G)+\lambda I$
\State $\hat P\gets(\hat G+\Lambda)^{-1}$, and $\hat B\gets-\mathcal H(\hat P)\diag(\hat P)^{-1}$
\State $\hat A\gets -\mathcal H(\hat P)$
\State $\hat V\gets$ the $K$ leading eigenvectors of $\hat A$, as columns
\State $\hat\beta\gets\operatorname{ppSPA}(\hat V,K)$ 
\State $\hat\beta_{:,k}\gets\hat\beta_{:,k}/\lVert\hat\beta_{:,k}\rVert_1$ for $k=1,\dots,K$
\end{algorithmic}
\end{algorithm}

\section{Experiments}
\label{sec:experiments}

\subsection{Recovery on synthetic data}
We perform a simulation by drawing $\beta\in\R_{+}^{m\times K}$ for $m=400$ items and $K=5$ topics: The first $K$ rows of $\beta$ form the identity matrix, and the remaining rows are drawn independently from the Dirichlet distribution with parameter $(0.3,\dots,0.3)$. The user factors $f_u$ are drawn independently from the Dirichlet distribution with parameter $2\gamma$, where $\gamma_k\propto3^{-(k-1)/(K-1)}$ and $\sum_k\gamma_k=1$. The interactions are drawn independently as $X_{ui}\sim\mathrm{Bernoulli}(0.1\,\langle\beta_{i,:},f_u\rangle)$. We fit EASE with $\lambda=250$ on $G=X^\top X$ for $n$ ranging from $500$ to $10^7$. Here, $\hat V$ is computed from the matrix $\hat A$ in \eqref{eq:observedA} with unit weights, and with $\Lambda=\lambda I$. With the sampled $\beta$, we compute $G^\star$ and $S^\star$ in \cref{lem:recovery} using \eqref{eq:gauge}. Since the size of $S^\star$ changes with $n$, we evaluate how far $\col(\hat{V})$ is from $\col(\beta_\Lambda)$ using the \emph{relative error}:
\begin{equation}\label{eq:rel}
    \min_{C \in \R^{K\times m}}\lVert S^\star-\hat VC\rVert/\lVert S^\star\rVert.
\end{equation}

\begin{figure}[t]
  \centering
  \includegraphics[width=\columnwidth]{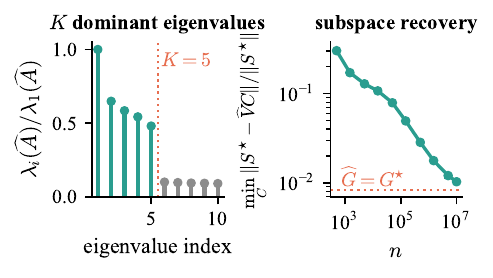}
  \Description{Two panels on synthetic data with K=5 topics. The left panel shows the 10 largest eigenvalues of the matrix A-hat, divided by the largest, with a visible cut-off after the fifth. The right panel shows the relative error of the estimated eigenspace, which decreases from 0.30 to 0.010 as the number of users grows from 500 to ten million. A horizontal dotted line at 0.0083 shows the limit of the relative error of the eigenspace computed from the expected co-occurrence matrix.}
  \caption{Subspace recovery on synthetic data. Left: the largest eigenvalues of $\hat A$ at $n=150000$. Right: the relative error $\min_C\lVert S^\star-\hat VC\rVert/\lVert S^\star\rVert$ against $n$, with the mean and one standard deviation over four seeds (the error bars are too small to be visible). The dotted line is the limit of the relative error with $\hat G=G^\star$ as $n\to\infty$.}
  \label{fig:synthrec}
\end{figure}

\Cref{fig:synthrec} shows the simulation results. The left panel displays the largest eigenvalues of $\hat A$ from one simulation with $n=150000$. We observe that the eigenvalues show a visible cut-off after the $K$-th largest one. This cut-off is consistent with \cref{lem:recovery}, by which $\hat A$ differs from the rank-$K$ matrix $S^\star$ by the diagonal matrix $\diag(S^\star)$ and the sampling noise in $\hat G$.

For the recovery of the subspace $\col(\beta_\Lambda)$, the right panel of \Cref{fig:synthrec} illustrates that the relative error \eqref{eq:rel} decreases as $n$ increases. This decrease is consistent with the first term of \eqref{eq:subspacebound}, which arises from the sampling noise in $\hat G$ and is expected to approach zero as $n\to\infty$. In contrast, the second term in \eqref{eq:subspacebound}, which arises from omitting the diagonal entries of $S^\star$, is irreducible as $n$ increases. We can see from the plot that the relative error approaches the irreducible error as $n \to \infty$.

\begin{table*}[t]
\centering
\caption{NDCG@10, recall@10 (Rec), and topic quality (TQ). Bold NDCG@10, and recall@10 numbers are based on paired two-sided $t$-tests over the users at 5\% level, and bold TQ numbers are based on bootstrap resampling tests at 5\% level.}
\label{tab:headline}
\resizebox{\linewidth}{!}{%
\setlength{\tabcolsep}{3pt}
\begin{tabular}{l*{18}{c}}
\toprule
& \multicolumn{3}{c}{\textbf{books}} & \multicolumn{3}{c}{\textbf{cds}} & \multicolumn{3}{c}{\textbf{gowalla}} & \multicolumn{3}{c}{\textbf{ml-1m}} & \multicolumn{3}{c}{\textbf{retail}} & \multicolumn{3}{c}{\textbf{steam}} \\
\cmidrule(lr){2-4}\cmidrule(lr){5-7}\cmidrule(lr){8-10}\cmidrule(lr){11-13}\cmidrule(lr){14-16}\cmidrule(lr){17-19}
\textbf{Model} & NDCG & Rec & TQ & NDCG & Rec & TQ & NDCG & Rec & TQ & NDCG & Rec & TQ & NDCG & Rec & TQ & NDCG & Rec & TQ \\
\midrule
WMF & 0.040 & 0.077 & $-$ & 0.046 & 0.085 & $-$ & 0.083 & 0.151 & $-$ & 0.029 & 0.051 & $-$ & 0.043 & 0.049 & $-$ & 0.018 & 0.033 & $-$ \\
LightGCN & 0.021 & 0.039 & $-$ & 0.017 & 0.032 & $-$ & 0.052 & 0.091 & $-$ & 0.025 & 0.042 & $-$ & 0.027 & 0.030 & $-$ & 0.027 & 0.054 & $-$ \\
SLIMElastic & \textbf{0.047} & \textbf{0.082} & $-$ & \textbf{0.051} & \textbf{0.092} & $-$ & 0.094 & 0.164 & $-$ & \textbf{0.042} & \textbf{0.072} & $-$ & \textbf{0.058} & \textbf{0.064} & $-$ & 0.028 & 0.057 & $-$ \\
ADMMSLIM & \textbf{0.046} & \textbf{0.081} & $-$ & \textbf{0.052} & \textbf{0.093} & $-$ & 0.096 & 0.168 & $-$ & 0.041 & 0.070 & $-$ & 0.056 & 0.060 & $-$ & 0.026 & 0.051 & $-$ \\
NCEPLRec & 0.040 & 0.073 & $-$ & \textbf{0.050} & \textbf{0.090} & $-$ & 0.093 & 0.167 & $-$ & \textbf{0.042} & \textbf{0.072} & $-$ & 0.056 & 0.062 & $-$ & 0.025 & 0.048 & $-$ \\
RecVAE & 0.039 & 0.072 & $-$ & 0.046 & 0.086 & $-$ & 0.089 & 0.165 & $-$ & 0.037 & 0.063 & $-$ & 0.046 & 0.050 & $-$ & 0.025 & 0.048 & $-$ \\
LDA & 0.018 & 0.038 & 0.140 & 0.014 & 0.029 & 0.129 & 0.073 & 0.135 & 0.300 & 0.030 & 0.052 & 0.313 & 0.031 & 0.033 & 0.297 & 0.023 & 0.046 & 0.179 \\
ETM & 0.018 & 0.038 & 0.137 & 0.007 & 0.014 & 0.024 & 0.032 & 0.065 & 0.004 & 0.025 & 0.042 & 0.060 & 0.031 & 0.032 & 0.115 & 0.027 & 0.055 & 0.046 \\
ECRTM & 0.020 & 0.044 & 0.199 & 0.033 & 0.061 & 0.141 & 0.043 & 0.068 & 0.104 & 0.015 & 0.027 & 0.155 & 0.020 & 0.020 & 0.185 & 0.007 & 0.014 & 0.028 \\
Mult-VAE & 0.030 & 0.058 & 0.064 & 0.033 & 0.063 & 0.039 & 0.092 & 0.164 & 0.090 & 0.025 & 0.041 & 0.092 & 0.030 & 0.031 & 0.106 & 0.025 & 0.050 & 0.069 \\
\midrule
NMF & 0.019 & 0.042 & 0.169 & 0.010 & 0.023 & 0.285 & 0.079 & 0.137 & 0.322 & 0.030 & 0.050 & 0.287 & 0.032 & 0.034 & 0.273 & \textbf{0.031} & \textbf{0.065} & 0.139 \\
NMF-ppSPA & 0.006 & 0.012 & 0.187 & 0.003 & 0.008 & 0.166 & 0.043 & 0.077 & 0.094 & 0.023 & 0.037 & 0.422 & 0.024 & 0.026 & 0.346 & \textbf{0.032} & \textbf{0.066} & 0.117 \\
\midrule
EASE & 0.045 & \textbf{0.081} & 0.191 & \textbf{0.051} & \textbf{0.092} & 0.219 & 0.093 & 0.163 & \textbf{0.498} & \textbf{0.043} & \textbf{0.073} & \textbf{0.495} & \textbf{0.059} & \textbf{0.066} & \textbf{0.459} & \textbf{0.029} & \textbf{0.061} & 0.130 \\
RW & 0.036 & 0.065 & 0.209 & 0.043 & 0.080 & 0.255 & \textbf{0.104} & \textbf{0.184} & \textbf{0.488} & 0.030 & 0.051 & 0.407 & 0.038 & 0.043 & 0.278 & 0.026 & 0.049 & 0.188 \\
Symmetric & 0.038 & 0.068 & \textbf{0.330} & 0.038 & 0.071 & 0.272 & 0.086 & 0.162 & \textbf{0.483} & 0.033 & 0.055 & 0.479 & 0.048 & 0.052 & 0.312 & 0.017 & 0.032 & 0.231 \\
DAN & \textbf{0.046} & \textbf{0.085} & 0.046 & 0.048 & 0.090 & 0.233 & \textbf{0.104} & \textbf{0.184} & \textbf{0.488} & \textbf{0.041} & 0.067 & 0.418 & 0.056 & 0.061 & 0.285 & \textbf{0.031} & \textbf{0.062} & 0.037 \\
RDLAE & \textbf{0.046} & \textbf{0.085} & 0.068 & \textbf{0.051} & \textbf{0.092} & 0.180 & 0.099 & \textbf{0.176} & 0.379 & \textbf{0.043} & \textbf{0.075} & 0.446 & \textbf{0.058} & \textbf{0.064} & 0.420 & \textbf{0.029} & \textbf{0.059} & 0.128 \\
\midrule
TEA & \textbf{0.046} & \textbf{0.083} & \textbf{0.331} & \textbf{0.051} & \textbf{0.096} & \textbf{0.380} & \textbf{0.100} & 0.176 & \textbf{0.508} & 0.040 & 0.068 & \textbf{0.492} & 0.056 & 0.062 & \textbf{0.472} & \textbf{0.029} & 0.058 & \textbf{0.290} \\
\bottomrule
\end{tabular}}
\end{table*}
\begin{table*}[t]
\centering
\footnotesize
\caption{Example topics recovered by TEA, each with its five highest-weight items. $n^{\mathrm{eff}}$ is the effective number of users who support the topic (defined in \cref{app:topics}). We omit \texttt{gowalla}, whose items are location identifiers without text.}
\label{tab:topics}
\begin{tabular}{@{}p{3.4cm}p{12.6cm}r@{}}
\toprule
\textbf{Topic} & \textbf{Top items} & $n^{\mathrm{eff}}$ \\
\midrule
\multicolumn{3}{@{}l}{\texttt{amazon-books}: five of the $K=6$ topics. Items are books.} \\
\midrule
young adult fiction & Mockingjay (The Hunger Games), Divergent, Catching Fire (Hunger Games, Book Two), The Hunger Games (Book 1), City of Bones (The Mortal Instruments, Book 1) & $22.5$ \\
Women's Murder Club series & 4th of July (Women's Murder Club), The 6th Target (Women's Murder Club), 2nd Chance (Women's Murder Club), 7th Heaven (Women's Murder Club), 10th Anniversary (Women's Murder Club) & $8.2$ \\
psychological thrillers & Don't Look for Me: A Novel, The Maid: A Novel (Molly the Maid Book 1), The Golden Couple: A Novel, The Plot: A Novel, Dark Roads: A Novel & $7.7$ \\
Nameless thriller series & In the Heart of the Fire (Nameless: Season One Book 1), The Mercy of Snakes (Nameless: Season One Book 5), The Praying Mantis Bride (Nameless: Season One Book 3), Red Rain (Nameless: Season One Book 4), Photographing the Dead (Nameless: Season One Book 2) & $16.7$ \\
physics workbooks & Physics Concept Questions - Book 1 (Mechanics 1), Physics Concept Questions - Book 3 (Matter), Physics Concept Questions - Book 4 (Heat), Physics Concept Questions - Book 2 (Mechanics 2), Physics Concept Questions - Book 5 (Sound) & $13.4$ \\
\midrule
\multicolumn{3}{@{}l}{\texttt{amazon-cds}: five of the $K=7$ topics. Items are albums.} \\
\midrule
Metallica & Master of Puppets, Ride the Lightning, Kill 'Em All, ...And Justice For All, Metallica & $32.6$ \\
KISS & Rock And Roll Over, Dynasty, Dressed To Kill, Creatures Of The Night, Kiss & $12.7$ \\
classic rock & Revolver, Abbey Road, The Beatles (The White Album), Meddle, Sgt. Pepper's Lonely Hearts Club Band & $35.0$ \\
mid-2000s pop & Love. Angel. Music. Baby., Breakaway, Autobiography, Speak, Rebirth & $18.7$ \\
hip-hop & Stillmatic, The College Dropout, The Black Album, Illmatic, The Blueprint & $14.1$ \\
\midrule
\multicolumn{3}{@{}l}{\texttt{ml-1m}: five of the $K=18$ topics. Items are films.} \\
\midrule
Disney animation & Cinderella, Sleeping Beauty, Dumbo, Peter Pan, Snow White and the Seven Dwarfs & $435.4$ \\
1990s science-fiction action & Judge Dredd, Johnny Mnemonic, Spawn, Universal Soldier, Timecop & $379.5$ \\
1980s comedy & Weird Science, Weekend at Bernie's, Police Academy, European Vacation, The Money Pit & $290.3$ \\
Clint Eastwood westerns & High Plains Drifter, The Outlaw Josey Wales, For a Few Dollars More, A Fistful of Dollars, Hang 'em High & $258.3$ \\
1990s horror & Scream 2, I Know What You Did Last Summer, Urban Legend, Thinner, Candyman & $206.0$ \\
\midrule
\multicolumn{3}{@{}l}{\texttt{online-retail-ii}: five of the $K=6$ topics. Items are stock lines.} \\
\midrule
polka-dot tableware & Red Retrospot Cup, Blue Polkadot Cup, Blue Polkadot Plate, Red Retrospot Plate, Pink Polkadot Plate & $152.9$ \\
printed bags & Jumbo Bag Vintage Doily, Lunch Bag Vintage Doily, Jumbo Bag Vintage Leaf, Lunch Bag Apple Design, Jumbo Bag Alphabet & $196.5$ \\
metal signs & Hand Over The Chocolate Sign, Gin + Tonic Diet Metal Sign, Please One Person Metal Sign, Cook With Wine Metal Sign, No Singing Metal Sign & $272.2$ \\
tea sets and retro kitchenware & Green Regency Teacup And Saucer, Roses Regency Teacup And Saucer, Pink Regency Teacup And Saucer, Regency Tea Plate Green, Regency Tea Plate Roses & $184.8$ \\
birthday cards and gift wrap & Spaceboy Birthday Card, Elephant Birthday Card, Tea Party Birthday Card, Cowboys And Indians Birthday Card, Spaceboy Gift Wrap & $173.5$ \\
\midrule
\multicolumn{3}{@{}l}{\texttt{steam}: the $K=5$ topics. Items are games.} \\
\midrule
Half-Life games & Half-Life 2: Episode One, Half-Life: Opposing Force, Half-Life: Blue Shift, Half-Life: Source, Half-Life 2: Lost Coast & $19.5$ \\
online multiplayer games & Trove, Brawlhalla, Garry's Mod, Team Fortress 2, BLOCKADE 3D & $69.4$ \\
anime JRPGs & Hyperdimension Neptunia Re;Birth1, Hyperdimension Neptunia Re;Birth2, Hyperdimension Neptunia Re;Birth3 V Generation, Hyperdimension Neptunia U: Action Unleashed, Fairy Fencer F & $10.6$ \\
strategy \& simulation & Stellaris, Sid Meier's Civilization: Beyond Earth, Crusader Kings II, Project Zomboid, Kerbal Space Program & $33.5$ \\
hand-drawn indie games & Nihilumbra, The Bridge, Dust: An Elysian Tail, Ballpoint Universe - Infinite, Botanicula & $9.2$ \\
\bottomrule
\end{tabular}
\end{table*}

\subsection{Benchmark on real datasets}

Our goal is to evaluate the recommendation performances and topic qualities of TEA on several public datasets. We detail our experiment below.

\paragraph{Datasets.}
We evaluate on six datasets: \texttt{amazon-books}, \texttt{amazon-cds} \citep{hou2024amazon},
\texttt{gowalla} \citep{cho2011friendship}, \texttt{ml-1m} \citep{harper2015movielens},
\texttt{online-retail-ii} \citep{chen2012online}, and \texttt{steam} \citep{kang2018sasrec}. For each dataset, we keep the $5000$ items with the most interactions and a random sample of at most $5000$ users. We split each user's interactions by time, holding out the items of the last timestamp for testing and those of the second-to-last timestamp for validation, without the items that the user interacted with earlier.

\paragraph{Baselines.} We compare TEA against several baselines, including six collaborative-filtering models: WMF \citep{hu2008implicit}, LightGCN \citep{he2020lightgcn}, SLIMElastic \citep{ning2011slim}, ADMMSLIM \citep{steck2020admm}, NCEPLRec \citep{wu2019nceplrec}, and RecVAE \citep{shenbin2020recvae}, and six topic models, LDA \citep{blei2003lda}, ETM \citep{dieng2020etm}, ECRTM \citep{wu2023ecrtm}, Mult-VAE \citep{liang2018multvae}, NMF of $G$ (NMF), and NMF of $\mathcal H(G)$, followed by ppSPA (NMF-ppSPA) \citep{arora2012topic}. We also compare TEA against five closed-form linear autoencoders: EASE \citep{steck2019ease}, RDLAE \citep{moon2023rlae}, DAN \citep{park2025dan}, and EASE with the random-walk (RW) and symmetric normalizations (Symmetric), which are commonly used for recommendation models \citep{choi2023bspm,cooper2014p3alpha,paudel2015rp3beta,Wang2019ngcf}.

\paragraph{Selection of $K$ and of the hyperparameters.}

For hyperparameter selection, we split the users into three sets. For each candidate value of $K$ and of the hyperparameters, we fit one model on the training items of each two of the three sets of users and evaluate the model on the validation items of those users. To select $K$, we extract the topics $\hat\beta$ from EASE with $\lambda=250$ by the last five steps of \cref{alg:tea}, and use them to predict each user's validation items. We model each validation item of user $u$ as a draw from the distribution $\sum_k\theta_{uk}\bar\beta_{:,k}$ over the items, where the weights $\theta_{uk}$ are fitted to the training items of $u$ and $\bar\beta_{:,k}=0.99\,\hat\beta_{:,k}+0.01\,\mathbf 1/m$ so that no item has probability zero. We score each $K$ by the log-perplexity over the validation items. We compute this score for every $K$ in $\{5,\dots,10,12,15,18,20,30,50,100,150,200,300\}$ and select the $K$ with the smallest mean log-perplexity. The selected $K$ is $6$ on
\texttt{amazon-books}, $7$ on \texttt{amazon-cds}, $30$ on \texttt{gowalla}, $18$ on \texttt{ml-1m},
$6$ on \texttt{online-retail-ii}, and $5$ on \texttt{steam}. We select the hyperparameters of every method on each dataset by NDCG@10 on the validation items. The selected values of the hyperparameters for all models are reported in \Cref{app:hyper}.

\paragraph{Evaluation.}
Each model is fitted with its selected hyperparameters on the training items of all users. The fitted models are then used to rank, for each user, all but their training and validation items. We report NDCG@10 and recall@10 of these rankings on the test items. To evaluate the topics $\hat\beta$ of each method, we use the \emph{topic quality} (TQ) \citep{dieng2020etm}, defined as $\mathrm{TD} \times \mathrm{TC}$, where the \emph{topic diversity} (TD) is the number of distinct items among the top $25$ items of the $K$ columns of $\hat\beta$, divided by $25K$, and the \emph{topic coherence} (TC) is
\begin{equation*}
  \mathrm{TC}=\frac{1}{K}\sum_{k=1}^{K}\frac{1}{45}\sum_{\{i,j\}\subset T_k}\frac{\log\big(\eta_{ij}/(\eta_i\eta_j)\big)}{-\log\eta_{ij}},
\end{equation*}
where $T_k$ is the top ten items of $\hat\beta_{:,k}$, $\eta_i$ is the fraction of users whose training items include item $i$, and $\eta_{ij}$ is the fraction of users whose training items include both $i$ and $j$ (the summand is $0$ if $\eta_{ij}=0$). Here, each summand is referred to as the \emph{normalized pointwise mutual information} (NPMI) between item $i$ and $j$ \citep{bouma2009normalized,aletras2013evaluating}. To compute $\mathrm{TQ}$, we hold out one third of the users, refit each model on the training items of the other two thirds, and then compute TQ on the held-out users. 

To compare NDCG@10 and recall@10 between any two methods, we employ paired two-sided $t$-tests across the test users at the $5\%$ level. Since TQ is not an average over users, we instead compare TQ using paired bootstrap resampling over the held-out users \citep{koehn2004statistical}.

\paragraph{Quantitative analysis of the results.}
\Cref{tab:headline} reports the three metrics of all the methods over the six datasets. The table shows that (1) TEA has the highest topic quality, or statistically ties with the highest topic quality, on all six datasets, and is significantly higher than every other method on \texttt{cds} and \texttt{steam}; (2) TEA's topic quality is significantly higher than that of every topic model on all six datasets; and (3) TEA has significantly higher topic quality than EASE on three datasets and statistically ties with EASE on the other three, which demonstrates that the normalizations and the dropout regularization that TEA adds to EASE improve the topic quality.

In terms of recommendation performance, linear autoencoders are as accurate as the collaborative filtering baselines: on every dataset, the best linear autoencoder is at most $0.001$ below the best baseline in NDCG@10. TEA is also competitive with the best method. In terms of NDCG@10, TEA statistically ties with the best method on \texttt{books}, \texttt{cds}, \texttt{gowalla}, and \texttt{steam}, and is at most $0.004$ below the best method on \texttt{ml-1m} and \texttt{retail}. In terms of recall@10, TEA is within $0.01$ of the best method on every dataset.

\paragraph{Qualitative analysis of the topics.}
\Cref{tab:topics} shows five topics of TEA for each dataset, except \texttt{gowalla} whose items are location identifiers. These topics range from broad genres to very narrow groups of items. For example, some topics consist of the albums of a single artist, such as KISS on \texttt{amazon-cds}, or the books of a single series, such as the Nameless series on \texttt{amazon-books}. These narrow topics appear because many users interacted with several albums or books of the same artist or series. Indeed, the effective number of users $n^{\mathrm{eff}}$ (defined in \cref{app:topics}) is above ten for both KISS and the Nameless series. Other topics group the items by both genre and period, such as 1980s comedy and 1990s horror on \texttt{ml-1m} and mid-2000s pop on \texttt{amazon-cds}. On \texttt{online-retail-ii}, the topics are product lines that share a common design, such as polka-dot tableware and Regency teacups and saucers, which suggests that the customers tend to buy matching items together.

{\emergencystretch=1em The effective number of users also helps explain the difference in topic quality between TEA and EASE. On \texttt{amazon-books}, \texttt{amazon-cds}, and \texttt{steam}, where TEA has significantly higher topic quality than EASE, the topics of EASE are supported by far fewer users than those of TEA. Hence, these topics of EASE can be dominated by the co-occurrences of a few very active users. On the other three datasets, the topics of EASE are already supported by many users, which may explain why the topic quality of TEA is not significantly higher than that of EASE on these datasets. This pattern suggests that the user normalization, which down-weights very active users, improves the topics when their co-occurrences come from a few users.\par}

\begin{figure}[!t]
  \centering
  \includegraphics[width=\columnwidth]{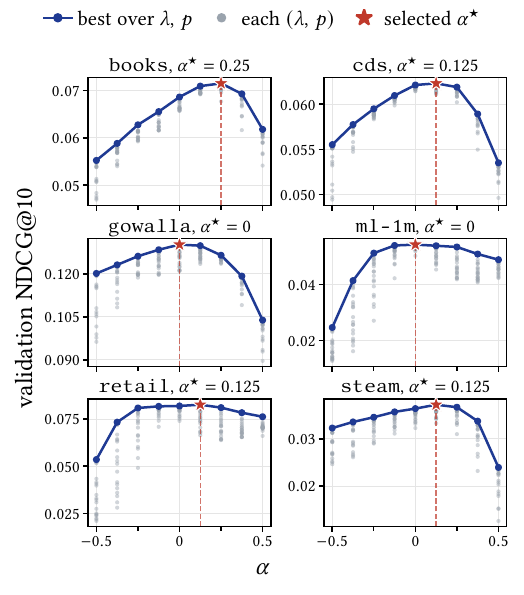}
  \Description{Six panels in three rows of two, one per dataset, plotting the validation NDCG@10 of TEA against the item exponent alpha from -0.5 to 0.5. Each panel shows the largest value over the regularization parameter and the dropout probability as a line, the value of each of their 20 pairs as dots, and the selected alpha as a star, whose value is in the panel title. On every dataset, the line rises up to the selected alpha, which is between 0 and 0.25, and falls toward alpha equal to -0.5 and 0.5.}
  \caption{Validation NDCG@10 of TEA against the item exponent $\alpha$. For each $\alpha$, the line is the largest value over the $20$ pairs of $\lambda$ and $p$, and the dots are the individual pairs.}
  \label{fig:alpha}
\end{figure}

\paragraph{Sensitivity to the item exponent.}
{\emergencystretch=1em To see how the item normalization affects the recommendations, we plot the validation NDCG@10 of TEA against the item normalization parameter $\alpha$ in \Cref{fig:alpha}. On all six datasets, the curve attains its maximum at an $\alpha\in\{0,0.125,0.25\}$, that is, when TEA does not reweight the items or slightly down-weights the popular items. As $\alpha$ moves away from this range, the NDCG@10 decreases, by up to $55\%$ at $\alpha=-0.5$ on \texttt{ml-1m} and $36\%$ at $\alpha=0.5$ on \texttt{steam}. Thus, the choice of $\alpha$ matters for the recommendation performance of TEA.\par}

\section{Related Work}
\label{sec:related}

\xhdr{User and item normalization.} Random-walk \citep{cooper2014p3alpha,paudel2015rp3beta} and symmetric \citep{Wang2019ngcf,he2020lightgcn,choi2023bspm} normalizations give the item-item matrices with entries $\sum_u X_{ui}X_{uj}/(g_id_u)$ and $\sum_u X_{ui}X_{uj}/(d_u\sqrt{g_ig_j})$, respectively. Other normalizations raise the degrees to exponents and include both matrices as special cases \citep{Fu2022neighborhood,park2025dan}. DAN \citep{park2025dan} is the closest method to ours, since both perform item and user normalization. However, DAN normalizes $G_{ij}$ by $g_i^{1-\alpha}g_j^{\alpha}$, while ours normalizes $G_{ij}$ by $g_i^{\alpha}g_j^{\alpha}$. Moreover, DAN's user normalization requires tuning a hyperparameter, whereas ours does not require any.

\xhdr{Latent-factor models for recommendations.} Given the latent variables of the users and items, the likelihood of the interactions can be a mixture over discrete latent classes \citep{hofmann2004latent}, a Gaussian distribution \citep{mnih2007pmf}, or a multinomial distribution whose parameters are the output of a neural network \citep{liang2018multvae}. Other methods fit the interactions by a matrix factorization with a squared loss weighted by implicit feedback, without an explicit generative model \citep{hu2008implicit}. Our model assumes that the conditional mean of each interaction is linear in a $K$-dimensional latent vector and that the interactions of a user are conditionally independent across items; examples of models that satisfy these assumptions are aspect Bernoulli models \citep{bingham2009aspect} and Poisson models \citep{gopalan2014ctpf}.

\xhdr{Topic models in recommender systems.} Several recommenders fit a topic model to the texts or reviews of the items jointly with a factorization of the interactions \citep{wang2011ctr,mcauley2013hft,gopalan2014ctpf}. The topics of these methods are distributions over the words of item texts or reviews. In contrast, our topics are distributions over the items, and we estimate the topics from only the user-item interactions.

\xhdr{Topics and structure from recommenders.} Post-hoc methods describe each factor of a matrix factorization by a topic model of the item descriptions \citep{rossetti2013explaining}. Hierarchical models factorize the interactions through a learned hierarchy of user and item clusters \citep{sugahara2024hierarchical}. Both approaches need an additional model, either a topic model of the item text or a new hierarchical recommender. In contrast, we extract the topics directly from the weight matrix of the recommender, demonstrating that the recommender can also be a topic model.

\section{Limitations}
\label{sec:limitations}
Our uniqueness result (\cref{thm:sep}) relies on \cref{ass:sep}, which requires that every topic has an anchor item, that is, an item that belongs to none of the other topics. Future work could relax this assumption, extend our analysis to more general latent-factor models, explore other user and item normalizations, and extend the recommendation to sequential or online settings.

\section{Conclusion}
\label{sec:conclusion}
We show that, under a latent-factor model of the user-item interactions, the topics of the items can be estimated from the weight matrix of a closed-form linear autoencoder. Thus, such an autoencoder provides both item recommendations and a topic model. In turn, the topics offer a way to interpret the recommendations. Building on this result, we propose TEA, which adds an item normalization and a tuning-free user normalization to a linear autoencoder. Our experiments demonstrate that TEA achieves significantly higher topic quality than every dedicated topic model and matches or exceeds the best linear autoencoder on all six public datasets, while its recommendation performance is competitive with the best tuned recommender.

\begin{acks}
    This work is supported by Faculty of Science, Chiang Mai University.
\end{acks}

\bibliographystyle{ACM-Reference-Format}
\bibliography{main}

\appendix

\section{Proof of \texorpdfstring{\cref{lem:lowrank}}{Lemma \getrefnumber{lem:lowrank}}}

For any user $u$ and items $i\neq j$, the entries $\hat X_{ui}$ and $\hat X_{uj}$ are independent. Moreover, $\E[\hat X_{ui}]=\langle\beta_{i,:},f_u\rangle$ and $\E[\hat X_{ui}^2]=\mathrm{Var}(\hat X_{ui})+(\E\hat X_{ui})^2$. Summing over $u$ therefore gives:
\begin{align*}
\E[\hat G]_{ij}&=\sum_{u}\E[\hat X_{ui}]\,\E[\hat X_{uj}]=(\beta\Phi\beta^\top)_{ij}\quad(i\neq j),\\
\E[\hat G]_{ii}&=\sum_{u}\big(\mathrm{Var}(\hat X_{ui})+(\E\hat X_{ui})^2\big)=(\beta\Phi\beta^\top)_{ii}+D_{ii}.
\end{align*}
Since $\Phi=\sum_uf_uf_u^\top$, we have $\Phi\succeq0$, and therefore $L\succeq0$ with $\rank(L)\le\rank(\Phi)\le K$. \qed

\section{Proof of \texorpdfstring{\cref{prop:topicdir}}{Proposition \getrefnumber{prop:topicdir}}}

Fix $k$. For any $v\in\R^m$, the Cauchy--Schwarz inequality gives:
\begin{align*}
(v^\top\beta_{:,k})^2&=\big((D^{1/2}v)^\top(D^{-1/2}\beta_{:,k})\big)^2\\
&\le(v^\top D v)(\beta_{:,k}^\top D^{-1}\beta_{:,k}).
\end{align*}
Thus, $(v^\top\beta_{:,k})^2/(v^\top D v)\le\beta_{:,k}^\top D^{-1}\beta_{:,k}$ for every nonzero $v$. Equality holds if and only if $D^{1/2}v$ is proportional to $D^{-1/2}\beta_{:,k}$, that is, $v\propto D^{-1}\beta_{:,k}$. Since $\rank(\beta)=K$, the column $\beta_{:,k}\ge0$ is nonzero. Since $D$ is a diagonal matrix with positive diagonal entries, we have $\mathbf{1}^\top D^{-1}\beta_{:,k}>0$. The constraint $\sum_i v_i=1$ therefore fixes the proportionality constant and gives the unique maximizer $t_k$.
\qed

\section{Proof of \texorpdfstring{\cref{lem:recovery}}{Lemma \getrefnumber{lem:recovery}}}
\label{app:recoveryproof}

Write $D_\Lambda=D+\Lambda$. By \cref{ass:lrd}, $D$ is a diagonal matrix with positive diagonal entries. Since the diagonal entries of $\Lambda$ are also positive, $D_\Lambda\succ0$. Since $\Phi\succ0$ and $\beta^\top\beta_\Lambda=\beta^\top D_\Lambda^{-1}\beta\succeq0$, we have $M\succ0$. Consequently, the Woodbury identity applied to $G^\star+\Lambda=D_\Lambda+\beta\Phi\beta^\top$ gives:
\begin{equation}\label{eq:woodbury}
P^\star=D_\Lambda^{-1}-S^\star.
\end{equation}
Since $M\succ0$, we can write $S^\star=\Gamma\Gamma^\top$ with $\Gamma=\beta_\Lambda M^{-1/2}$, so $S^\star\succeq0$. Since $M^{-1/2}$ is invertible, $\rank(S^\star)=\rank(\beta_\Lambda)$ and $\col(S^\star)=\col(\beta_\Lambda)$. Since $\rank(\beta)=K$ and $D_\Lambda^{-1}$ is invertible, $\rank(\beta_\Lambda)=K$. Finally, $D_\Lambda^{-1}$ is a diagonal matrix, so $-\mathcal H(P^\star)=\mathcal H(S^\star)$. \qed

\section{Proof of \texorpdfstring{\cref{prop:ident}}{Proposition \getrefnumber{prop:ident}}}
\label{app:identproof}

As in the proof of \cref{lem:recovery}, write $S^\star=\Gamma\Gamma^\top$ with $\Gamma=\beta_\Lambda M^{-1/2}$. By \cref{lem:recovery}, $P^\star_{ab}=-S^\star_{ab}$ for all items $a\neq b$.

We first show that, for any disjoint sets $I,J\subset\{1,\dots,m\}$ of $K$ items each, $P^\star_{I,J}$ is invertible if and only if $\beta_{I,:}$ and $\beta_{J,:}$ are invertible. The proof of \cref{lem:recovery} shows that $D_\Lambda$ is a diagonal matrix with positive diagonal entries and that $M\succ0$. Consequently, the $K\times K$ matrices $(D_\Lambda^{-1})_{I,I}$ and $M^{-1/2}$ are invertible. Since $D_\Lambda^{-1}$ is a diagonal matrix, $\Gamma_{I,:}=(D_\Lambda^{-1})_{I,I}\,\beta_{I,:}\,M^{-1/2}$. Hence, the matrix $\Gamma_{I,:}$ is invertible if and only if $\beta_{I,:}$ is invertible. The same holds for $\Gamma_{J,:}$ and $\beta_{J,:}$. Moreover, $S^\star_{I,J}=\Gamma_{I,:}\Gamma_{J,:}^\top$, so $\det S^\star_{I,J}=\det \Gamma_{I,:}\,\det \Gamma_{J,:}$. Since $I\cap J=\emptyset$, every pair $(a,b)\in I\times J$ has $a\neq b$. Hence, $P^\star_{I,J}=-S^\star_{I,J}$. Therefore, $P^\star_{I,J}$ is invertible if and only if $\beta_{I,:}$ and $\beta_{J,:}$ are invertible.

Next, we prove \eqref{eq:sstar}. Fix an item $i$, and let $I$ and $J$ be as in \cref{ass:ident}. Since $\beta_{I,:}$ and $\beta_{J,:}$ are invertible, the matrices $\Gamma_{I,:}$, $\Gamma_{J,:}$, and $S^\star_{I,J}=-P^\star_{I,J}$ are invertible. Since $S^\star_{i,J}=\Gamma_{i,:}\Gamma_{J,:}^\top$, $S^\star_{I,J}=\Gamma_{I,:}\Gamma_{J,:}^\top$, and $S^\star_{I,i}=\Gamma_{I,:}\Gamma_{i,:}^\top$, we obtain:
\begin{equation}\label{eq:schurid}
S^\star_{i,J}(S^\star_{I,J})^{-1}S^\star_{I,i}=\Gamma_{i,:}\Gamma_{J,:}^\top\big(\Gamma_{J,:}^\top\big)^{-1}\Gamma_{I,:}^{-1}\Gamma_{I,:}\Gamma_{i,:}^\top=\Gamma_{i,:}\Gamma_{i,:}^\top=S^\star_{ii}.
\end{equation}
Since $i\notin I\cup J$ and $I\cap J=\emptyset$, every pair of indices $(a,b)$ in $\{i\}\times J$, $I\times J$, or $I\times\{i\}$ has $a\neq b$. Hence, $S^\star_{i,J}=-P^\star_{i,J}$, $S^\star_{I,J}=-P^\star_{I,J}$, and $S^\star_{I,i}=-P^\star_{I,i}$. Substituting these three equations into $S^\star_{ii}=S^\star_{i,J}(S^\star_{I,J})^{-1}S^\star_{I,i}$ gives:
\[
S^\star_{ii}=-P^\star_{i,J}(P^\star_{I,J})^{-1}P^\star_{I,i},
\]
which is \eqref{eq:sstar}.

It remains to prove the uniqueness of $S^\star$. Let $S'\in\R^{m\times m}$ have rank at most $K$ and satisfy $\mathcal H(S')=-\mathcal H(P^\star)=\mathcal H(S^\star)$, where the second equality holds by \cref{lem:recovery}. Fix an item $i$, and let $I$ and $J$ be as in \cref{ass:ident}. Consider the submatrix $S'_{\{i\}\cup I,\{i\}\cup J}$, whose entries are $S'_{ab}$ with $a\in\{i\}\cup I$ and $b\in\{i\}\cup J$. Since $i\notin I\cup J$ and $I\cap J=\emptyset$, the only index in both $\{i\}\cup I$ and $\{i\}\cup J$ is $i$. Hence, $a\neq b$ for every such pair $(a,b)$ other than $(i,i)$, so $S'_{ab}=S^\star_{ab}$ because $\mathcal H(S')=\mathcal H(S^\star)$. Consequently, with $i$ listed first among the rows and among the columns of this submatrix,
\[
S'_{\{i\}\cup I,\{i\}\cup J}=\begin{pmatrix}S'_{ii}&S^\star_{i,J}\\ S^\star_{I,i}&S^\star_{I,J}\end{pmatrix}.
\]
Since $I$ and $J$ have $K$ items each and $i\notin I\cup J$, this submatrix is $(K+1)\times(K+1)$. Since the rank of a submatrix is at most the rank of the matrix, this submatrix has rank at most $\rank(S')\le K$ and is therefore singular. Moreover, $S^\star_{I,J}$ is invertible, since $\beta_{I,:}$ and $\beta_{J,:}$ are invertible. Consequently, the determinant formula for the Schur complement gives:
\[
0=\det S'_{\{i\}\cup I,\{i\}\cup J}=\det(S^\star_{I,J})\big(S'_{ii}-S^\star_{i,J}(S^\star_{I,J})^{-1}S^\star_{I,i}\big).
\]
Since $\det(S^\star_{I,J})\neq0$, this equation and \eqref{eq:schurid} give:
\[
S'_{ii}=S^\star_{i,J}(S^\star_{I,J})^{-1}S^\star_{I,i}=S^\star_{ii}.
\] Since the item $i$ is arbitrary, $\diag(S')=\diag(S^\star)$. Combined with $\mathcal H(S')=\mathcal H(S^\star)$, this gives $S'=S^\star$. Finally, by \cref{lem:recovery}, $S^\star$ has rank $K$ and $\mathcal H(S^\star)=-\mathcal H(P^\star)$. Hence, $S^\star$ is the only matrix of rank at most $K$ whose off-diagonal matrix is $-\mathcal H(P^\star)$. \qed

\section{Proof of \texorpdfstring{\cref{thm:sep}}{Theorem \getrefnumber{thm:sep}}}
\label{app:sepproof}

We prove \cref{thm:sep}. The notation is that of \cref{sec:identification}, with $D_\Lambda=D+\Lambda$.

\begin{proof}[Proof of \cref{thm:sep}]
Since $\rank(\beta)=K$ and $D_\Lambda^{-1}$ is invertible, $\beta_\Lambda$ has rank $K$. Let the columns of $O\in\R^{m\times K}$ form an orthonormal basis of $\col(\beta_\Lambda)$. Since $OO^\top$ is the orthogonal projection onto $\col(\beta_\Lambda)$, we have $\beta_\Lambda=OO^\top\beta_\Lambda=ON$ with $N:=O^\top\beta_\Lambda\in\R^{K\times K}$. Since $K=\rank(\beta_\Lambda)=\rank(ON)\le\rank(N)$, the matrix $N$ is invertible. Write $o_i^\top$ and $r_i^\top$ for the $i$-th rows of $O$ and $\beta_\Lambda$. Since $\beta_\Lambda=ON$, we have $r_i^\top=o_i^\top N$, that is, $o_i^\top=r_i^\top N^{-1}$. Let $c_k^\top$ be the $k$-th row of $N^{-1}$. Consequently, the vectors $c_1,\dots,c_K$ are linearly independent.

We claim that $\beta_\Lambda$ is separable. Since the entries of both $\beta$ and $D_\Lambda^{-1}$ are non-negative, the entries of $\beta_\Lambda$ are non-negative. Moreover, each row of $\beta_\Lambda$ has zero entries in the same positions as the same row of $\beta$. Consequently, every anchor of $\beta$ is an anchor of $\beta_\Lambda$. Since the $K$ anchors of $\beta$ are distinct items, $\beta_\Lambda$ is separable.

Now, we show that every separable non-negative basis of $\col(\beta_\Lambda)$ is $\beta_\Lambda$ up to permutation and positive column scaling. Since $r_i\ge0$, each $o_i=\sum_k(r_i)_k\,c_k$ lies in the cone $\mathcal C := \operatorname{cone}\{c_1,\dots,c_K\}$, that is, the set of non-negative linear combinations of $c_1,\ldots,c_k$. By the definition of $o_i$, it is immediate that $\operatorname{cone}\{o_1,\dots,o_m\}\subseteq\mathcal C$. Conversely, since $\beta_\Lambda$ is separable, each column $k$ of $\beta_\Lambda$ has an anchor $a_k$, so $r_{a_k}=\theta_ke_k$ for some $\theta_k>0$, where $e_k$ is the $k$-th standard basis vector of $\R^K$. Hence, $c_k=o_{a_k}/\theta_k$ for every $k$. Therefore, $\operatorname{cone}\{o_1,\dots,o_m\}\supseteq\mathcal C$, and so $\operatorname{cone}\{o_1,\dots,o_m\}=\mathcal C$. 

Now let $W'\in\R_+^{m\times K}$ be a separable matrix whose columns form a basis of $\col(\beta_\Lambda)$. As for $\beta_\Lambda$, we have $W':=ON'$ for some invertible matrix $N'$. Let ${c'_k}^\top$ be the $k$-th row of ${N'}^{-1}$. Since $W'$ is non-negative and separable, the same argument as for $\beta_\Lambda$ gives: 
\[\operatorname{cone}\{c'_1,\dots,c'_K\}=\operatorname{cone}\{o_1,\dots,o_m\}=\mathcal C.\]
Consequently, $\{c_1,\dots,c_K\}$ and $\{c'_1,\dots,c'_K\}$ are linearly independent sets that generate the same cone $\mathcal C$. Hence, these two (ordered) sets of vectors differ by a permutation and positive scaling. Since the rows of $N^{-1}$ and ${N'}^{-1}$ are these two sets of vectors, there are a permutation matrix $\Pi$ and a diagonal matrix $\Upsilon$ with positive diagonal entries such that ${N'}^{-1}=\Upsilon\Pi^\top N^{-1}$. Since $\Pi^{-1}=\Pi^\top$, we have
\[
W'=ON'=ON\Pi\Upsilon^{-1}=\beta_\Lambda\Pi\Upsilon^{-1},
\]
as claimed.
\end{proof}

\section{Proof of \texorpdfstring{\cref{thm:subspace}}{Theorem \getrefnumber{thm:subspace}}}
\label{app:subspaceproof}

We use the notation of \cref{sec:identification,sec:observed}, with $D_\Lambda=D+\Lambda$. Write:
\[
\Psi=P^\star-\hat P,\qquad\delta=\lVert\hat G-G^\star\rVert,\qquad\Delta=\lVert\hat A-S^\star\rVert,
\]
and let $\lambda_j(\cdot)$ denote the $j$-th largest eigenvalue. By \cref{lem:recovery} and \eqref{eq:woodbury}, $P^\star=D_\Lambda^{-1}-S^\star$, $S^\star\succeq0$, and $\rank(S^\star)=K$. Since $\hat P=P^\star-\Psi$, $\mathcal H$ is linear, and $\mathcal H(D_\Lambda^{-1})=0$, \eqref{eq:observedA} gives:
\begin{equation}\label{eq:Adecomp}
\hat A=-\mathcal H(P^\star)+\mathcal H(\Psi)=S^\star-\diag(S^\star)+\mathcal H(\Psi).
\end{equation}
Moreover, since $G^\star+\Lambda=L+D_\Lambda\succeq D_\Lambda\succeq\nu I$, the matrix $P^\star = (G^\star + \Lambda)^{-1}$ is positive definite and $\lVert P^\star\rVert\le1/\nu$. Hence, $0\preceq S^\star\preceq D_\Lambda^{-1}$ and $\lVert S^\star\rVert\le1/\nu$.

We claim that
\begin{equation}\label{eq:twoDelta}
\min_{C\in\R^{K\times m}}\lVert S^\star-\hat VC\rVert\le2\Delta.
\end{equation}
Taking $C=\hat V^\top S^\star$ gives $\min_C\lVert S^\star-\hat VC\rVert\le\lVert(I-\hat V\hat V^\top)S^\star\rVert$. Since the columns of $\hat V$ are orthonormal eigenvectors of $\hat A$ associated with the eigenvalues $\lambda_1(\hat A),\dots,\lambda_K(\hat A)$, we have $\lVert(I-\hat V\hat V^\top)\hat A\rVert=\max_{j>K}|\lambda_j(\hat A)|$. Since $S^\star\succeq0$ has rank $K$, $\lambda_j(S^\star)=0$ for every $j>K$, so Weyl's inequality gives $|\lambda_j(\hat A)|\le\Delta$ for every $j>K$. Consequently, since $\lVert I-\hat V\hat V^\top\rVert\le1$, the triangle inequality gives:
\[
\lVert(I-\hat V\hat V^\top)S^\star\rVert\le\lVert(I-\hat V\hat V^\top)\hat A\rVert+\lVert(I-\hat V\hat V^\top)(\hat A-S^\star)\rVert\le2\Delta.
\]
Moreover, by \eqref{eq:Adecomp}, $\hat A-S^\star=\mathcal H(\Psi)-\diag(S^\star)$. Since the diagonal entries of $S^\star\succeq0$ are non-negative, the triangle inequality gives $\Delta\le\lVert\mathcal H(\Psi)\rVert+\max_iS^\star_{ii}$.

Now, we show that $\max_iS^\star_{ii}\le\rho/\nu$. Fix an item $i$. Since $G^\star+\Lambda\succ0$, the Cauchy--Schwarz inequality gives $P^\star_{ii}\,(G^\star_{ii}+\Lambda_{ii})\ge1$. Since $G^\star_{ii}+\Lambda_{ii}-(D_\Lambda)_{ii}=L_{ii}$, $(D_\Lambda)_{ii}\ge\nu$, and $\Lambda_{ii}>0$, \eqref{eq:woodbury} gives:
\begin{align}
S^\star_{ii}&=\frac{1}{(D_\Lambda)_{ii}}-P^\star_{ii}\le\frac{1}{(D_\Lambda)_{ii}}-\frac{1}{G^\star_{ii}+\Lambda_{ii}}\nonumber\\
&=\frac{L_{ii}}{(D_\Lambda)_{ii}(G^\star_{ii}+\Lambda_{ii})}\le\frac{L_{ii}}{\nu\,G^\star_{ii}}\le\frac{\rho}{\nu}.\label{eq:Sdiag}
\end{align}

Finally, we prove \eqref{eq:subspacebound} in two regimes of $\delta$:
\begin{itemize}
\item If $\delta\le\nu/5$, we first bound $\lVert\mathcal H(\Psi)\rVert$. Since $\mathcal H(\Psi)=\Psi-\diag(\Psi)$ and $\max_i|\Psi_{ii}|\le\lVert\Psi\rVert$, we have $\lVert\mathcal H(\Psi)\rVert\le2\lVert\Psi\rVert$. By Weyl's inequality, $\lambda_m(\hat G+\Lambda)\ge\lambda_m(G^\star+\Lambda)-\delta\ge\nu-\delta\ge4\nu/5$, so $\lVert\hat P\rVert\le5/(4\nu)$. Since $\Psi=P^\star\big((\hat G+\Lambda)-(G^\star+\Lambda)\big)\hat P=P^\star(\hat G-G^\star)\hat P$, we obtain $\lVert\mathcal H(\Psi)\rVert\le2\lVert P^\star\rVert\,\delta\,\lVert\hat P\rVert\le5\delta/(2\nu^2)$. Combining this bound with \eqref{eq:twoDelta}, $\Delta\le\lVert\mathcal H(\Psi)\rVert+\max_iS^\star_{ii}$, and \eqref{eq:Sdiag} gives:
\[
\min_C\lVert S^\star-\hat VC\rVert\le2\lVert\mathcal H(\Psi)\rVert+2\max_iS^\star_{ii}\le\frac{5\delta}{\nu^2}+\frac{2\rho}{\nu}.
\]
\item If $\delta>\nu/5$, then $5\delta/\nu^2>1/\nu$. Taking $C=0$ and using $\rho\ge0$, we obtain:
\[
\min_C\lVert S^\star-\hat VC\rVert\le\lVert S^\star\rVert\le\frac1\nu<\frac{5\delta}{\nu^2}\le\frac{5\delta}{\nu^2}+\frac{2\rho}{\nu}.
\]
\end{itemize}
Hence, \eqref{eq:subspacebound} holds in both regimes. \qed

\begin{table*}[t]
\caption{Selected hyperparameters of each method on each dataset.}
\label{tab:hyper}
\fontsize{8}{9.5}\selectfont
\renewcommand{\arraystretch}{0.92}
\begin{tabular}{ll*{6}{r}}
\toprule
\textbf{Model} & \textbf{Hyperparameter} & \textbf{books} & \textbf{cds} & \textbf{gowalla} & \textbf{ml-1m} & \textbf{retail} & \textbf{steam} \\
\midrule
WMF & number of factors & 128 & 128 & 64 & 128 & 128 & 64 \\
 & regularization & 0.01 & 0.01 & 0.01 & 0.01 & 0.01 & 0.01 \\
\addlinespace[1pt]
LightGCN & learning rate & 0.001 & 0.01 & 0.01 & 0.001 & 0.001 & 0.001 \\
 & number of layers & 2 & 2 & 2 & 2 & 3 & 3 \\
\addlinespace[1pt]
SLIMElastic & regularization & 0.01 & 0.01 & 0.01 & 0.1 & 0.1 & 0.1 \\
 & L1 ratio & 0.01 & 0.01 & 0.001 & 0.01 & 0.01 & 0.001 \\
\addlinespace[1pt]
ADMMSLIM & L1 regularization & 0.5 & 0.5 & 0.5 & 0.5 & 5 & 1 \\
 & L2 regularization & 50 & 200 & 500 & 500 & 500 & 500 \\
\addlinespace[1pt]
NCEPLRec & regularization & 1000 & 1000 & 10000 & 10000 & 10000 & 10000 \\
 & rank & 400 & 400 & 400 & 400 & 400 & 400 \\
\addlinespace[1pt]
RecVAE & latent dimension & 200 & 200 & 200 & 200 & 200 & 200 \\
 & KL scale & 0.005 & 0.005 & 0.005 & 0.0025 & 0.0025 & 0.005 \\
\midrule
LDA & number of iterations & 20 & 100 & 20 & 100 & 20 & 20 \\
 & prior on topic proportions & 0.5 & 0.5 & 0.05 & 0.5 & 0.5 & 0.5 \\
 & prior on topics & $1/K$ & $1/K$ & 0.1 & 0.1 & $1/K$ & $1/K$ \\
\addlinespace[1pt]
ETM & learning rate & 0.01 & 0.005 & 0.01 & 0.005 & 0.005 & 0.002 \\
 & hidden size & 800 & 400 & 800 & 800 & 800 & 400 \\
 & dropout probability & 0.5 & 0 & 0 & 0 & 0 & 0.5 \\
 & KL weight & 0.2 & 0.5 & 0.1 & 0.1 & 0.5 & 0.5 \\
\addlinespace[1pt]
ECRTM & learning rate & 0.005 & 0.005 & 0.005 & 0.005 & 0.005 & 0.005 \\
 & temperature & 0.2 & 0.2 & 0.1 & 0.2 & 0.1 & 0.1 \\
 & ECR weight & 20 & 20 & 20 & 50 & 200 & 200 \\
 & KL weight & 0.2 & 0.1 & 0.2 & 0.5 & 0.2 & 0.5 \\
\addlinespace[1pt]
Mult-VAE & learning rate & 0.003 & 0.003 & 0.003 & 0.003 & 0.003 & 0.003 \\
 & KL weight & 0.2 & 0.1 & 0.5 & 0.5 & 0.5 & 0.5 \\
 & dropout probability & 0.5 & 0.5 & 0 & 0.5 & 0 & 0 \\
\midrule
NMF & regularization & 0 & 0 & 0.01 & 0.01 & 0.1 & 0.01 \\
 & L1 ratio & 0 & 0 & 1 & 1 & 1 & 0 \\
 & initialization & \texttt{nndsvda} & \texttt{nndsvd} & \texttt{nndsvd} & \texttt{nndsvd} & \texttt{nndsvd} & \texttt{nndsvd} \\
\addlinespace[1pt]
NMF-ppSPA & regularization & 0.01 & 0.01 & 0.1 & 0 & 0 & 0.1 \\
 & L1 ratio & 0.5 & 0.5 & 1 & 0 & 0 & 0 \\
 & initialization & \texttt{nndsvda} & \texttt{nndsvda} & \texttt{nndsvd} & \texttt{nndsvd} & \texttt{nndsvd} & \texttt{nndsvda} \\
\midrule
EASE & $\lambda$ & 50 & 50 & 100 & 1000 & 500 & 1000 \\
\addlinespace[1pt]
RW & $\lambda$ & 250 & 1000 & 50 & 50 & 50 & 500 \\
\addlinespace[1pt]
Symmetric & $\lambda$ & 1000 & 100 & 1000 & 50 & 50 & 100 \\
\addlinespace[1pt]
DAN & $\lambda$ & 1000 & 250 & 50 & 50 & 50 & 500 \\
 & $\alpha$ & 0.25 & 0.25 & 0 & 0.5 & 0.5 & 0.25 \\
 & $\beta$ (the user exponent) & 0.2 & 0.5 & 1 & 0 & 0 & 0.2 \\
\addlinespace[1pt]
RDLAE & $\lambda$ & 50 & 50 & 100 & 250 & 500 & 500 \\
 & $p$ & 0.5 & 0.3 & 0.9 & 0.3 & 0.1 & 0.7 \\
 & $\xi$ & 0.5 & 0.1 & 0.1 & 0.5 & 0.25 & 0.1 \\
\midrule
TEA & $\alpha$ & 0.25 & 0.125 & 0 & 0 & 0.125 & 0.125 \\
 & $p$ & 0.7 & 0.7 & 0.7 & 0.3 & 0 & 0.7 \\
 & $\lambda$ & 250 & 250 & 50 & 250 & 50 & 250 \\
\bottomrule
\end{tabular}
\end{table*}

\section{Proof of \texorpdfstring{\cref{thm:userweight}}{Theorem \getrefnumber{thm:userweight}}}
\label{app:userweight}

We first prove \eqref{eq:signal}. Since $\mathbf 1^\top x_ux_u^\top\mathbf 1=d_u^2$ and the diagonal entries of $x_ux_u^\top$ sum to $d_u$, we have $\mathbf 1^\top\mathcal H(x_ux_u^\top)\mathbf 1=d_u(d_u-1)$. Since $\bar G=\sum_uq_ux_ux_u^\top$ and the weights $q_u$ depend only on $d$, summing over the users gives \eqref{eq:signal}.

Next, we prove \eqref{eq:noise}. Here, $\mathrm{Var}(Y\mid d)=\E[Y^2\mid d]-\E[Y\mid d]^2=\E\big[(Y-\E[Y\mid d])^2\mid d\big]\succeq0$ for a symmetric random matrix $Y$. Since $x_u$ is binary, $(x_ux_u^\top)^2=(x_u^\top x_u)\,x_ux_u^\top=d_u\,x_ux_u^\top$. Hence,
\[
\mathrm{Var}(Y_u\mid d)=q_u^2\big(d_uA_u-A_u^2\big)\preceq q_u^2\,d_u\,A_u,
\]
where we used $A_u^2\succeq0$. Since $0\preceq\mathrm{Var}(Y_u\mid d)\preceq q_u^2d_uA_u$, we have $\lVert\mathrm{Var}(Y_u\mid d)\rVert\le q_u^2d_u\lVert A_u\rVert$ for every user $u$. Hence, the triangle inequality gives:
\[
\Big\lVert\sum_{u}\mathrm{Var}(Y_u\mid d)\Big\rVert\le\sum_{u}q_u^2d_u\lVert A_u\rVert=\sum_{u}q_u^2v_u,
\]
which is \eqref{eq:noise}.

Finally, we find the weights that minimize $F(q)$. A user with $d_u=0$ has $\sigma_u=v_u=0$ and does not affect $F(q)$. Every other user has $v_u>0$, since $A_u\succeq0$ and $\operatorname{tr}(A_u)=d_u>0$. Over these users, the Cauchy--Schwarz inequality gives:
\[
\begin{aligned}
  \sum_{u} q_u\sigma_u&=\sum_{u}(q_u\sqrt{v_u})(\sigma_u/\sqrt{v_u})\\
  &\le\Big(\sum_{u} q_u^2v_u\Big)^{1/2}\Big(\sum_{u}\sigma_u^2/v_u\Big)^{1/2}.
\end{aligned}
\]
Since $\sum_uq_u\sigma_u>0$, this gives $F(q)\ge(\sum_u\sigma_u^2/v_u)^{-1/2}$, with equality if and only if $q_u\sqrt{v_u}$ is proportional to $\sigma_u/\sqrt{v_u}$, that is, $q_u\propto\sigma_u/v_u=(d_u-1)/\lVert A_u\rVert$. \qed

\section{Hyperparameters}
\label{app:hyper}

We select the hyperparameters of every method on each dataset by NDCG@10 on the validation items. Mult-VAE \citep{liang2018multvae}, ETM \citep{dieng2020etm}, and ECRTM \citep{wu2023ecrtm} also keep the training epoch with the highest validation NDCG@10. \Cref{tab:hyper} reports the selected hyperparameters of every method. For the linear autoencoders, $\lambda$ is the regularization parameter, $\alpha$ is the item exponent, and $p$ is the dropout probability. For RDLAE, $\xi$ is the bound on the diagonal entries of $B$.

\section{Selection of the example topics}
\label{app:topics}

\balance

Among all the topics, we choose to display in \cref{tab:topics} the ones with both high coherence and a large effective number of users, defined below. Specifically, let $\hat\beta$ be the topics of TEA fitted on the training items of all users, let $T_k$ be the top ten items of $\hat\beta_{:,k}$, and let $\mathrm{TC}_k$ be the average NPMI over the pairs in $T_k$, with $\eta_i$ and $\eta_{ij}$ taken over all users. For each user $u$, let $\omega_{uk}$ be the fraction of the co-occurrences of pairs of items in $T_k$, counted over the training items of all users, that come from user $u$. The \emph{effective number of users} of topic $k$ is the reciprocal Simpson index $n^{\mathrm{eff}}_k=(\sum_u\omega_{uk}^2)^{-1}$, which equals $1$ when a single user provides all the co-occurrences and equals $r$ when $r$ users provide equal shares. We display the five topics with the largest values of the following score:
\[
  \mathrm{TC}_k\cdot\frac{\log n^{\mathrm{eff}}_k}{\log n}.
\]
Here, the factor $\log n^{\mathrm{eff}}_k/\log n$ is $0$ when a single user provides all the co-occurrences and $1$ when all $n$ users provide equal shares.

\end{document}